\documentclass[11pt]{article}
\usepackage{amsfonts}
\usepackage{graphicx}
\usepackage{amsmath}
\usepackage{amssymb}
\usepackage{latexsym}
\usepackage{color}
\usepackage{indentfirst}
\usepackage{cite}
\usepackage{float}
\usepackage{subcaption}

\begin{document}

	\thispagestyle{empty}
	
	\begin{center}
		\vspace{1.8cm}
		{\Large \textbf{
				Asymmetric EPR Steering in a Cavity-Magnon System Generated by a Squeezed Vacuum Field and an Optical Parametric Amplifier
			}}


		\vspace{1.5cm}
		\renewcommand{\thefootnote}{\alph{footnote}}
		\textbf{Abdelkader Hidki$^{\textbf{}}$,}$^{1,}${\footnote{email: abdelkader.hidki$@$gmail.com}}  \textbf{Noureddine Benrass,}$^{1}$\,
	\textbf{Abderrahim Lakhfif,}$^{2}$
		\, \textbf{and} \textbf{Mostafa Nassik}$^{1}$
		\vspace{0.5cm}
		
		$^{1}$\textit{LPTHE, Department of Physics, Faculty of Sciences, Ibn Zohr University, Agadir, Morocco}\\[0.5em]
		$^{2}$\textit{LPHE-Modeling and Simulation, Faculty of Sciences, Mohammed V University in Rabat, Rabat, Morocco}\\[0.5em]
		
		\vspace{3cm}\textbf{Abstract}
	\end{center}

We investigate a cavity-magnon system with two magnon modes coupled to a common cavity microwave field. The cavity is integrated with an optical parametric amplifier (OPA) and driven by a squeezed vacuum field. The introduction of the OPA and the squeezed vacuum field induce squeezing in the cavity mode, which is  transferred to the magnon modes through magnetic dipole interactions. Our findings demonstrate that enhancing the OPA gain and the squeezing parameter significantly enhances the quantum entanglement and the Einstein-Podolsky-Rosen (EPR) steering. Furthermore, the photon-magnon coupling strength can be adjusted to control the directionality of EPR steering, offering a mechanism for achieving one-way EPR steering under specific conditions. This control is fine-tuned by varying system parameters, thereby providing a robust platform for steering in the presence of thermal noise. Our findings advance the understanding of macroscopic quantum correlations and hold promising implications for quantum information processing, particularly in generating, manipulating, and enhancing quantum steering phenomena. This practical aspect of our research will inspire hope for future applications in the field of quantum information.\\


	\baselineskip=18pt \medskip
	\noindent  
	\newpage
	
	\section{Introduction}


Quantum correlations, characterized by unique properties, are typically classified into several types, including Bell nonlocality, quantum entanglement, and steering,  depending on different criteria \cite{reid1989demonstration,wiseman2007steering,jones2007entanglement,cavalcanti2009experimental}.  Entanglement is understood as the inseparability between two or more parties, serving as a crucial resource for processes such as quantum dense coding \cite{li2002quantum} and quantum teleportation \cite{braunstein1998teleportation}. Continuous variable entanglement, in particular, has garnered significant interest due to its relative simplicity and efficiency in  manipulation, generation, and detection \cite{braunstein2005quantum}. The concept of steering, initially introduced by Schrödinger in response to the Einstein-Podolsky-Rosen (EPR) paradox \cite{einstein1935can}, describes a situation where one party can influence the state of another via local measurements. Recently, Wiseman et al., in 2007, formalized the concept of EPR steering by relating it to the violation of a local hidden state model \cite{wiseman2007steering,jones2007entanglement}. EPR steering can manifest as  one-way (asymmetric) or two-way (symmetric). However, two-way steering has received less focus since its properties resemble those of quantum entanglement despite differing underlying conditions. Both theoretical and experimental research on EPR steering has been extensive, as it offers significant applications in areas like one-sided device-independent quantum cryptography \cite{gehring2015implementation,branciard2012one}, secure quantum teleportation \cite{he2015secure}, and subchannel discrimination \cite{reid2013signifying}.

	Recently, ferrimagnetic materials, particularly yttrium-iron-garnet (YIG) spheres, have attracted substantial attention owing to their high spin density and low dissipation rates. Unlike traditional optomechanical systems \cite{aspelmeyer2014cavity,kibret2024generation,amazioug2023strong,kibret2024generation2}, hybrid quantum systems involving magnons exhibit unique characteristics \cite{lachance2019hybrid}. 	One significant mode in these systems is the Kittel mode, corresponding  to a uniform spin wave mode in a YIG sphere. This mode can exhibit strong coupling with microwave photons, forming cavity-magnon polaritons \cite{huebl2013high,zhang2014strongly}. Furthermore, the Kittel mode \cite{kittel1948theory} can interact with various systems involving  discrete  or continuous variables \cite{li2018magnon,zhang2016cavity,sohail2023entanglement,ren2022long,ren2022chiral,hidki2024generation,xie2023nonreciprocal,sohail2023distant,tadesse2024distant,hidki2024entanglement,sohail2023enhanced}. The coherent interaction between spin waves and microwave photons is  vital to developing sophisticated hybrid systems, making it a key focus of current research. Magnons--excitations of spin-wave modes--can coherently couple with microwave photons via magnetic dipole interactions, forming cavity magnonics \cite{zhang2014strongly}. Simultaneously, they can also couple with a mechanical mode via magnetostrictive interaction, which involves the vibrations of the YIG sphere, forming cavity magnomechanics \cite{zhang2016cavity}. These interactions are crucial for developing new technologies in these fields. 	 	The photon-magnon system serves as an up-and-coming platform for realizing coherent optical phenomena. This includes  magnetically controllable slow light \cite{kong2019magnetically}, magnomechanically induced transparency \cite{zhang2016cavity}, magnon-polariton bistability \cite{wang2018bistability}, nonreciprocal microwave field transmission \cite{wang2019nonreciprocity}, and high-order sideband generation \cite{liu2018magnon}. Additionally, this hybrid system with quantum magnonics has facilitated the exploration of numerous intriguing quantum information tasks, such as magnon–magnon entanglement and steering  \cite{zheng2021enhanced}, photon-magnon–phonon entanglement \cite{li2018magnon}, feedback control of quantum correlations  \cite{amazioug2023feedback}, magnon cooling  \cite{sharma2018optical}, steady Bell-state generation  \cite{yuan2020steady}, enhanced quantum entanglement by an optical parametric amplifier (OPA) \cite{hidki2023enhanced1,lakhfif2024maximum,hidki2024entanglement2},  magnon blockade  \cite{liu2019magnon}, and entropy production rate and correlations \cite{edet2024entropy}.
	 	
	 In this paper, we explore a cavity magnonic system composed of two YIG spheres inside a microwave cavity field. Previous studies on similar systems have led to remarkable advancements in cavity magnonics, including magnon entanglement induced by the Kerr effect \cite{zhang2019quantum}, dissipative coupling between magnon and photon modes \cite{wang2019nonreciprocity, xu2019cavity}, long-distance coupling of YIG spheres \cite{zare2018indirect}, and the discovery of higher-order exceptional points \cite{zhang2019higher}. Other notable achievements include controlling one-way quantum steering and generating entanglement using a squeezed vacuum field produced by a Josephson parametric amplifier (JPA) \cite{nair2020deterministic, yang2021controlling}, and quantum correlations facilitated by an OPA and parametric one-way quantum steering through control induced by four-wave mixing \cite{noura2024enhanced,wang2023parametric}. Here, we propose a novel approach distinct from previous methods. We investigate a cavity-magnon system with a JPA and an OPA.  We aim to demonstrate the feasibility of achieving controllable one-way EPR steering alongside maximal entanglement. In this context, "controllable" signifies the ability to achieve  either one-way steering ($1 \rightarrow 2$ and  $2 \rightarrow 1$) or two-way EPR steering ($1 \leftrightarrow 2$) by fine-tuning various system parameters.

The structure of this paper is as follows: Section \ref{sec2} outlines the model and Hamiltonian of the cavity-magnon system. Section \ref{sec3} delves into the detailed examination of the steady-state solution, entanglement, and steering properties. In Section \ref{sec4}, we present the numerical results and our analysis. Finally, we draw our conclusions in Section \ref{sec5}.

	\section{Physical system}\label{sec2}
	\begin{figure}[H]
		\centerline{\includegraphics[width=12cm]{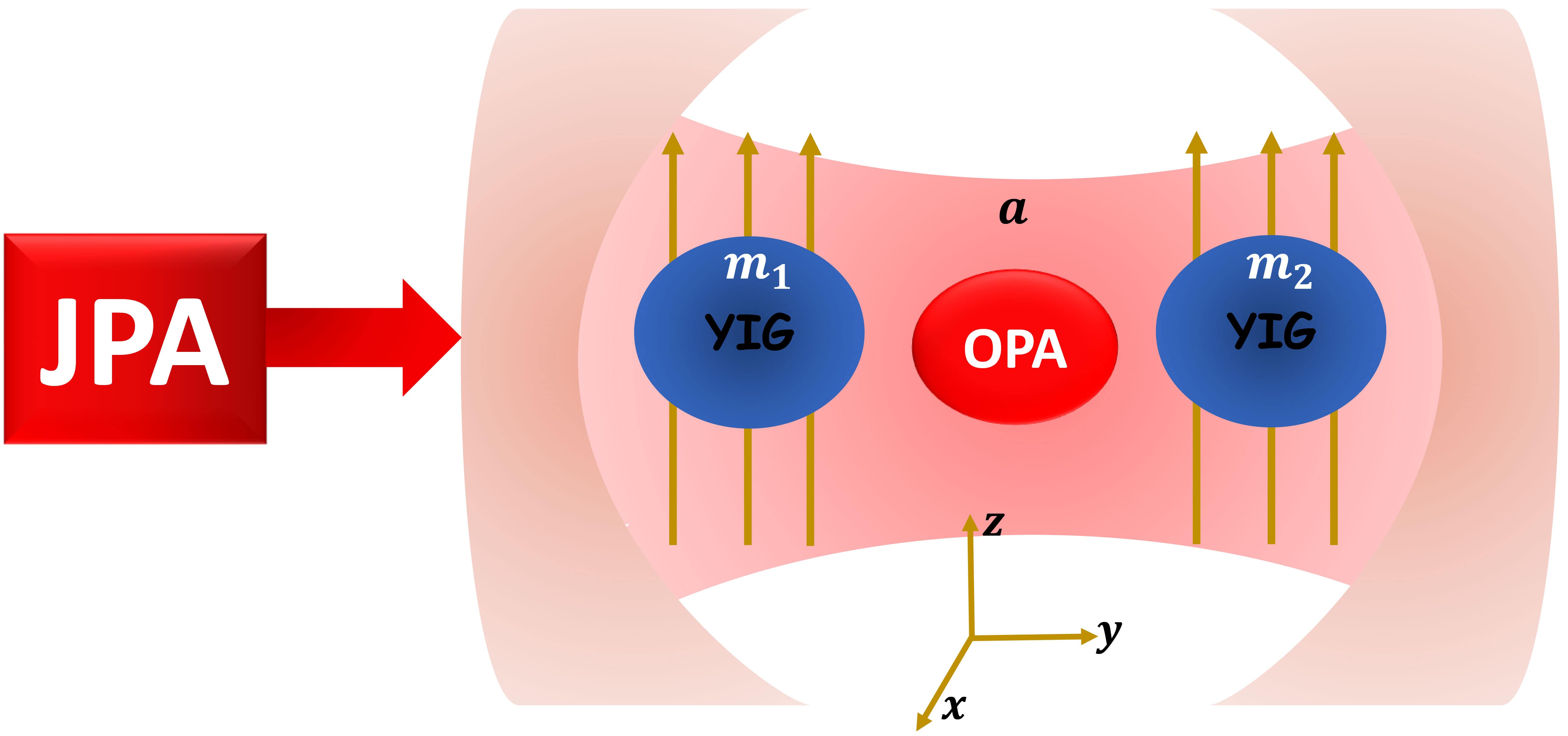}}
		\caption{\footnotesize{Diagram of the cavity-magnon system. Two YIG spheres hosting a magnon mode are positioned within a microwave cavity. The spheres are placed close to the magnetic field of the cavity mode along the $x$-axis and simultaneously within a bias magnetic field along the $z$-axis. The cavity field contains an OPA and is also driven by a JPA.
		}}
		\label{model}
	\end{figure}

We examine a hybrid cavity-magnon system \cite{yang2021controlling,nair2020deterministic} that includes a microwave cavity mode and two magnon modes, as depicted in Fig. \ref{model} (a). Magnons are quasiparticles representing a collective excitation of numerous spins, each with a spin value $5/2$ within a YIG sphere. These magnons interact with a single cavity photon through magnetic dipole coupling. The Hamiltonian of the system in the frame rotating at  the squeezed vacuum field $\omega_s$ is given by: 
	\begin{eqnarray}\label{E1}
	\mathcal{H} =\Delta _{a}a^{\dagger }a+\sum_{k=1,2}\left[ \Delta_{k}m_{k}^{\dagger }m_{k}+\Gamma_{k}(a\,m_{k}^{\dagger }+a^{\dagger}m_{k})\right] +i\Lambda(a^{\dagger 2} e^{i\varphi}-a^{2} e^{-i\varphi}),
 \end{eqnarray}
where  $a$ and $a^\dagger $ denote the annihilation and creation operators of the cavity mode, respectively, where $[a,a^{\dagger}]=1 $, similarly, $m_1$ and $m_2$ ($m_1^\dagger$ and $m_2^\dagger$) are the annihilation  (creation) operators for the two magnon modes, where $[m_k,m_k^{\dagger}]=1 $ ($k=1,2$). The magnon modes are used to describe the collective motion of spins, and this description is made possible by the Holstein–Primakoff transformation \cite{holstein1940field}, which expresses these modes in terms of bosonic operators. The parameters $\Delta_a = \omega_a - \omega_s$ and $\Delta_k = \omega_k - \omega_s$ ($k=1,2$) are, respectively,  the detunings of the cavity frequency $\omega_a$ and the $k^\text{th}$ magnon from the frequency $\omega_s$ of the squeezed vacuum field. The frequencies of the magnon modes are set by the bias magnetic fields  $B_j$ and can be expressed as $\omega_j = \gamma_0 B_j$, where  $\gamma_0 = 2\pi \times 28 \, \text{GHz/T}$ is the gyromagnetic ratio. Furthermore, the parameter $\Gamma_{k}$  denotes the coupling strength between the cavity and the $k^\text{th}$ magnon modes. This coupling can be tuned by altering the orientation of the bias magnetic fields and the placement of the YIG spheres. The last term represents the interaction between the cavity and the OPA, which is defined by the parametric gain $\Lambda$ and the phase $\varphi$.

Considering the effects of the decay rates of cavity and magnon modes, the dynamic behavior of the hybrid cavity-magnon system can be characterized by the following set of quantum Langevin equations (QLEs):
      \begin{align}
      	\dot{a}&=-(i\Delta_{a}+\kappa_a)a-i\Gamma_{1}m_1-i\Gamma_{2}m_2+2\Lambda a^{\dagger}e^{i\varphi}+\sqrt{2\kappa_a}a^{\text{in}},\\
      	\dot{m}_k&=-(i\Delta_{k}+\kappa_k)m_k-i\Gamma_{k}m_k+\sqrt{2\kappa_k}m_k^{\text{in}}, \,(k=1,2),
      \end{align}
where $\kappa_a$ and $\kappa_k$ represent the dissipation rates of the cavity mode and the $k^\text{th}$ magnon modes, respectively, while $a^{\text{in}}$ and $m^{\text{in}}_k$ denote the input noise operators for the cavity, the $k^\text{th}$ magnon mode, these noise operators have a zero mean value, and their correlation functions satisfy the following conditions \cite{gardiner2004quantum}:
	\begin{eqnarray}
	\left\langle a^{in}(t)a^{in\dagger }(t^{\prime }) ; a^{in\dagger }(t)a^{in}(t^{\prime })
	\right\rangle  &=&[\mathcal{N}+1;\mathcal{N}]\delta (t-t^{\prime }),\text{ \ }  \label{E4} \\
	\left\langle a^{in}(t)a^{in}(t^{\prime });a^{in\dagger }(t)a^{in\dagger }(t^{\prime })\right\rangle  &=&%
	[\mathcal{M};\mathcal{M}^{\ast}]\delta (t-t^{\prime }),\,  \label{E5} \\
	\left\langle m_{k}^{in}(t)m_{k}^{in\dagger }(t^{\prime })  ; m_{k}^{in\dagger }(t)m_{k}^{in}(t^{\prime
	}) \right\rangle 
	&=&[\bar{n}_{k}+1;\bar{n}_{k}]\delta (t-t^{\prime }), \, (k=1,2),\label{E6}
\end{eqnarray}%
where $\mathcal{N} = \bar{n}_a \cosh^2 r + (\bar{n}_a + 1) \sinh^2 r$, and $\mathcal{M} = (1 + 2\bar{n}_a) e^{i\theta} \cosh r \sinh r$. Here, $r$ is the squeezing parameter, and $\theta$ represents the phase of the squeezed vacuum field. Additionally, $\bar{n}_o=1/[\exp(\hbar\omega_o/k_B T) -1]$ ($o = a, 1, 2$), with $k_B$ as the Boltzmann constant and $T$ as the environment temperature.


\section{Steady-state solution, entanglement and steering}\label{sec3}

To analyze the entanglement and steering between the two magnon modes, we introduce the quadrature of the cavity field and the two magnon modes:
$  x_{a}=\frac{1}{\sqrt{2}}(a^{\dagger }+a),  y_{a}=\frac{i}{\sqrt{2}}( a^{\dagger }- a), X_{k}=\frac{1}{\sqrt{2}}(  m_{k}^{\dagger }+ m_{k}),  Y_{k}=\frac{i}{\sqrt{2}}(  m_{k}^{\dagger }- m_{k})  $ ($k=1,2$). The same definitions apply to the input noise operators. The QLEs governing the quadrature fluctuations ($x_{a}$, $y_{a}$, $X_{1}$, $Y_{1}$, $X_{2}$, $Y_{2}$) can be expressed in the following matrix form:
          \begin{equation}
          	\dot{v}=A\,v(t)+n(t),
          \end{equation}
where $v(t) = [x_a(t), y_a(t), X_1(t), Y_1(t), X_2(t), Y_2(t)]^\text{T}$ is the vector for the quantum fluctuations, and $n(t) = [\sqrt{2\kappa_a}x_a^{\text{in}}(t), \sqrt{2\kappa_a}y_a^{\text{in}}(t), \sqrt{2\kappa_1}X_1^{\text{in}}(t), \sqrt{2\kappa_1}Y_1^{\text{in}}(t), \sqrt{2\kappa_2}X_2^{\text{in}}(t), \sqrt{2\kappa_2}Y_2^{\text{in}}(t)]^\text{T}$ is the vector for noises. The drift matrix $A$ is then defined as follows:
         \begin{equation}
         	A =
         	\begin{pmatrix}
         		-\kappa_a+2\Lambda\cos(\varphi) & \Delta_a+2\Lambda\sin(\varphi) & 0 & \Gamma_1 & 0 & \Gamma_2 \\
         		-\Delta_a+2\Lambda\sin(\varphi) & -\kappa_a-2\Lambda\cos(\varphi) & -\Gamma_1 & 0 & -\Gamma_2 & 0 \\
         		0 & \Gamma_1 & -\kappa_1 & \Delta_1 & 0 & 0 \\
         		-\Gamma_1 & 0 & -\Delta_1 & -\kappa_1 & 0 & 0 \\
         		0 & \Gamma_2 & 0 & 0 & -\kappa_2 & \Delta_2 \\
         		-\Gamma_2 & 0 & 0 & 0 & -\Delta_2 & -\kappa_2
         	\end{pmatrix}.
         \end{equation}
The system is a three-mode Gaussian state in continuous variables, and it can be fully characterized in phase space by a $6\times6$ covariance matrix (CM)  $\Sigma$. This matrix is defined by the elements $\Sigma_{ij}(t) = \frac{1}{2}\langle v_i(t)v_j(t) + v_j(t)v_i(t) \rangle $ ($i,j=1,...,6$). The steady-state covariance matrix $\Sigma$ can be found by solving the Lyapunov equation \cite{vitali2007optomechanical}:
      \begin{equation}
      	A\,\Sigma+\Sigma \,A^{\text{T}}=-\mathcal{F},
      \end{equation}
where $\mathcal{F}$ is the diffusion matrix defined by the elements $\mathcal{F}_{ij}\delta(t-t') = \frac{1}{2}\langle n_i(t)n_j(t) + n_j(t)n_i(t) \rangle $ ($i,j=1,...,6$), and given explicitly by:
          \begin{equation}
        	\mathcal{F} =
        	\begin{pmatrix}
        		\alpha^{+} & \beta & 0 & 0 & 0 & 0 \\
        		\beta & \alpha^{-} & 0 & 0 & 0 & 0 \\
        		0 & 0 & \kappa_1 (2\bar{n}_{1}+1) & 0 & 0 & 0 \\
        	0 & 0 &0 & \kappa_1(2\bar{n}_{1}+1) & 0 & 0 \\
        		0 & 0 & 0 & 0 & \kappa_2(2\bar{n}_{2}+1) & 0\\
        		0 & 0 & 0 & 0 & 0 & \kappa_2 (2\bar{n}_{2}+1)
        	\end{pmatrix},
        \end{equation}
    with $\alpha^{+}=(\mathcal{M}+\mathcal{M}^{\ast}+2\mathcal{N}+1)\kappa_a$,  $\alpha^{-}=(-\mathcal{M}-\mathcal{M}^{\ast}+2\mathcal{N}+1)\kappa_a$ and $\beta=i(\mathcal{M}^{\ast}-\mathcal{M})\kappa_a$.

The level of squeezing is quantified by:
         \begin{equation}
         S_p = -10 \log_{10} \left( \frac{\langle p^2 \rangle}{\langle p^2 \rangle_{\text{zpf}}} \right),
         \end{equation}
  where $p$ represents either $X_k$ or $Y_k$ and $\langle p^2 \rangle_{\text{zpf}}$ refers to the zero-point fluctuation of  $p$. $S_p >0$ dB indicates that the associated mode is squeezed.

For two-mode Gaussian states of the magnons  $m_1$ and $m_2$, a practical criterion for Gaussian quantum steering, derived from the concept of quantum coherent information, has been established \cite{kogias2015quantification}. Regarding quantum entanglement, the logarithmic negativity $E_N$ is a convenient measure \cite{adesso2004extremal,vitali2007optomechanical}. All these metrics can be calculated using the reduced $4\times4$ CM $\Sigma_r$ for $m_1$ and $m_2$, which corresponds to the last four rows and columns of the CM $\Sigma$:
            \begin{equation}
         	\Sigma_r =
         	\begin{pmatrix}
         		\Sigma_1 & \Sigma_c  \\
         		\Sigma_c^{\text{T}} & \Sigma_2  \\
         	\end{pmatrix},
         \end{equation}
with $\Sigma_1$, $\Sigma_2$, and $\Sigma_c$  represent  $2\times2$ sub-matrices of the reduced CM $\Sigma_r$. In this context,  the logarithmic negativity($E_{12}$)  for the two magnons  is determined as follows \cite{adesso2004extremal}:
               \begin{equation}
             E_{12}=\max[0,-\ln(2\mu^{-})],
            \end{equation}
with,
\begin{equation}
	\mu^{-} \equiv 2^{-\frac{1}{2}} \left( \Delta - \sqrt{\Delta^2 - 4\det(\Sigma_r)} \right)^{\frac{1}{2}},
\end{equation}
and
\begin{equation}
	\Delta \equiv \det(\Sigma_1) + \det(\Sigma_2) - 2\det(\Sigma_c).
\end{equation}
Furthermore, the Gaussian quantum steering  is expressed as \cite{kogias2015quantification}:
     \begin{align}
     \mathcal{G}^{1 \to 2} (\Sigma_r) &= \max \left[ 0, \frac{1}{2} \ln \left( \frac{\det(\Sigma_1)}{4\det(\Sigma_r)} \right) \right],\\
     \mathcal{G}^{2 \to 1} (\Sigma_r) &= \max \left[ 0, \frac{1}{2} \ln \left( \frac{\det(\Sigma_2)}{4\det(\Sigma_r)} \right) \right].
     \end{align} 
To analyze the asymmetry in the steerability of the two-mode Gaussian state, we propose a measure of steering asymmetry defined as \cite{liao2020controlling}:
      \begin{equation}
      	\mathcal{G}^{S}=\left|\mathcal{G}^{1 \to 2} -\mathcal{G}^{2 \to 1}\right|.
      \end{equation}


\section{Numerical results and analysis}\label{sec4}

	\begin{figure}[tbh!]
	\centerline{\includegraphics[width=16cm]{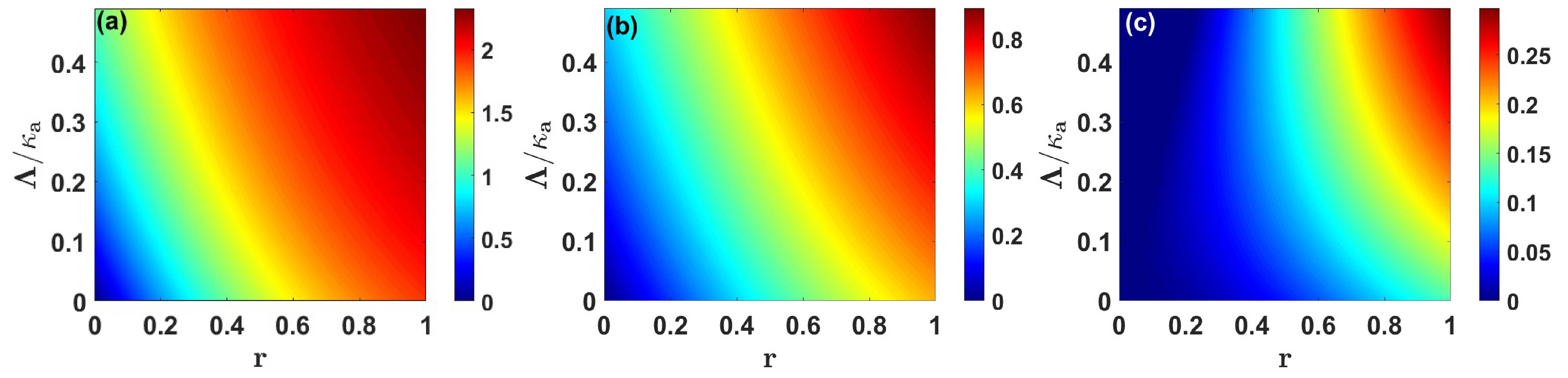}}
	\caption{\footnotesize{ (a) Squeezing metrics of the two magnons, (b) entanglement levels, and (c) degrees of steering  versus $r$ and $\Lambda/\kappa_a$.	The other parameters are given in the text.
	}}
	\label{F2}
\end{figure}

 In this section, we explore the steady-state entanglement ($ E_{12} $) and directional steering between the two magnon modes ($\mathcal{G}^{1\to2}$, $\mathcal{G}^{2\to1}$). For our analysis, we utilize the experimentally feasible parameters provided in Refs. \cite{yang2021controlling, li2018magnon}:\\
  $\omega_a/2\pi=\omega_1/2\pi=\omega_2/2\pi=10$ GHz, $\kappa_a/2\pi=5$ MHz,  $\kappa_1=\kappa_2=\kappa_a/5$, $\Gamma_{1}=\Gamma_{2}=4\kappa_a$, $T=20$ mK, $\theta=0$ and optimal detunings $\Delta_a=\Delta_1=\Delta_2=0$ \cite{nair2020deterministic}. After selecting these parameters, we determined that the maximum value of $\Lambda$, which meets the stability criteria, is $\Lambda=0.49 \kappa_a$, with the optimal phase being $\varphi=0$.

 To gain insight into the mechanisms responsible for generating   steering and entanglement, we analyze the   steering, squeezing, and entanglement levels.  Fig. \ref{F2} shows (a) the squeezing of two magnons, (b) the steady-state magnon–magnon entanglement $E_{12}$, and (c) the symmetric steering versus the squeezing parameter $r$ and the gain of the OPA $\Lambda/\kappa_a$. The consideration of a symmetric system leads to equal degrees of steering and magnon squeezing  ($\mathcal{G}^{1\to2} = \mathcal{G}^{2\to1}$ and $S_{ X_1} = S_{ X_2}$). Without the squeezed vacuum field, there is no steering; in contrast, the entanglement and squeezing of magnon modes can be generated by the OPA alone or by the squeezed vacuum field. As the value of $r$ increases, all the steering, squeezing, and entanglement can be generated, underscoring the essential role of the squeezed  field in creating quantum correlations. The squeezing of magnons increases with  $\Lambda$ and $r$, which enhances both entanglement and steering. 
 
 Furthermore, increasing the gain of the OPA improves both the degrees of steering and entanglement   produced by the squeezed  field. As a result, both the JPA and the OPA contribute to improving steering and entanglement   by squeezing the cavity field. The presence of the OPA and JPA leads to the squeezing of the cavity field \cite{li2019squeezed,hidki2023transfer,noura2024enhanced}. This squeezing is then transferred to the two magnon modes via linear beam-splitter photon-magnon interactions, which improves entanglement and steering between the magnon modes.

	\begin{figure}[tbh!]
	\centerline{\includegraphics[width=15cm]{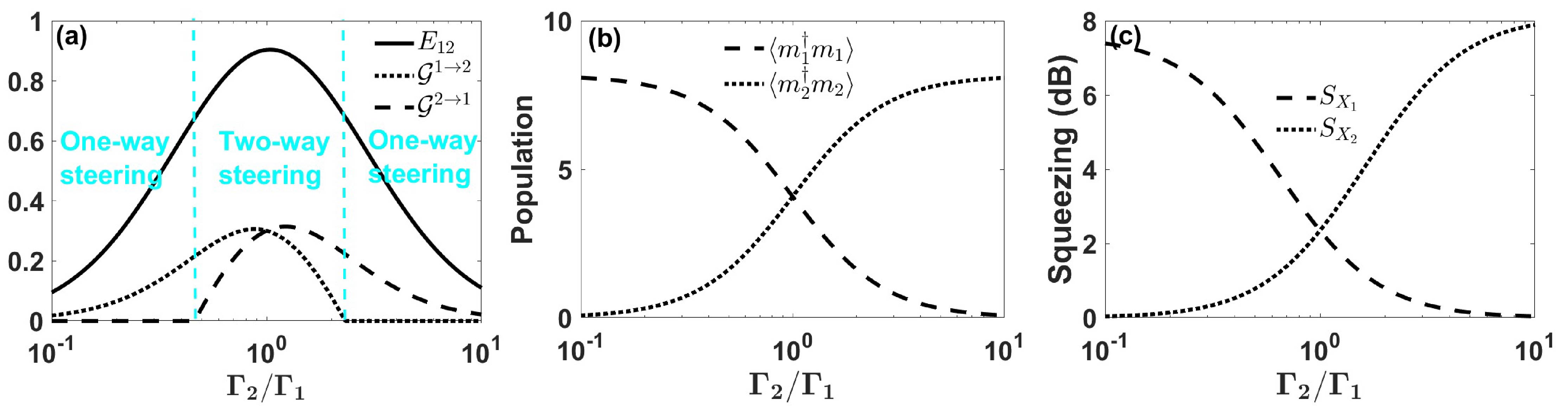}}
	\caption{\footnotesize{Plots of (a) magnon-magnon entanglement ($E_{12}$) and quantum steering ($\mathcal{G}^{1 \to 2}$ and $\mathcal{G}^{2 \to 1}$), (b) the population of the two magnons, and (c) the  magnon squeezing versus the ratio of  $\Gamma_{2}/\Gamma_{1}$.
	}}
	\label{F3}
\end{figure}

Next, we investigate the control of asymmetric steering in the system. Fig. \ref{F3}(a) shows how the magnon-magnon entanglement, denoted as $ E_{12}$ and the degree of quantum steering vary with respect to the ratio of the two photon-magnon coupling rates, $\Gamma_{2}/\Gamma_{1}$. A key observation is that the maximum entanglement between the magnon modes occurs when the system is symmetric, i.e., $\Gamma_{2} = \Gamma_{1}$, where the steering is equal in both directions. As the ratio $\Gamma_{2}/\Gamma_{1}$ deviates from $1$, creating an asymmetry in the system, the entanglement $E_{12}$ initially increases, suggesting that a slight imbalance in coupling strengths can enhance quantum correlations. However, with further deviation, $E_{12}$ gradually decreases, indicating that excessive asymmetry reduces the ability of the system to maintain strong entanglement. This phenomenon can be attributed to a balanced interaction optimally mediating quantum information between the modes. At the same time, a large disparity leads to one mode dominating the interaction, thereby reducing the overall entanglement. Interestingly, the directionality of steering also shifts with asymmetry. When $\Gamma_{2}/\Gamma_{1} > 1$, the steering  $\mathcal{G}^{2 \to 1}$ is more pronounced than the reverse $\mathcal{G}^{1 \to 2}$. Conversely, when $\Gamma_{2}/\Gamma_{1} < 1$, the steering $\mathcal{G}^{1 \to 2}$ becomes stronger than $\mathcal{G}^{2 \to 1}$. This indicates that the dominant coupling direction dictates the dominant steering direction, as stronger coupling to one mode enables more effective control and measurement of the quantum state of the other. Furthermore, the specific ratios at which the steering in each direction reaches its peak—approximately $\Gamma_{2}/\Gamma_{1} \approx 0.87$ for $\mathcal{G}^{1 \to 2}$ and $\Gamma_{2}/\Gamma_{1} \approx 1.27$ for $\mathcal{G}^{2 \to 1}$—highlight that there is an optimal level of asymmetry for maximizing steering in a particular direction. This control over the directionality of steering by tuning the photon-magnon coupling ratios provides a versatile tool for manipulating quantum correlations in the system. Such control is crucial for tasks in quantum information processing, where specific directional steering may be required to achieve desired outcomes, such as secure quantum communication.

To explore the potential relationship between quantum steering, magnon squeezing, and the population of magnon modes in the system, Figs. \ref{F3}(b) and \ref{F3}(c) present the population and the squeezing of the two magnons versus the ratio of the two photon-magnon coupling strengths, $\Gamma_{2}/\Gamma_{1}$. In Fig. \ref{F3}(b), as the coupling strength ratio increases, the population of mode $m_1$ decreases while the population of mode $m_2$ increases. At $\Gamma_{2}/\Gamma_{1} = 1$, the system's symmetrical structure results in equal populations for the two magnon modes. As the effective photon-magnon coupling strength $\Gamma_{2}$ ($\Gamma_{1}$) increases, the interaction between the cavity photons and the magnon in mode $m_2$ ($m_1$) becomes stronger, leading to a larger population in mode $m_2$ ($m_1$). This indicates that the mode with the stronger coupling interacts more intensely with the photons, absorbing more energy and thus increasing its population. Compared with the results in  Fig. \ref{F3}(a), it becomes evident that a magnon mode with a larger population is more challenging to steer by the other mode. This observation aligns with the findings in Ref. \cite{zheng2019manipulation}, which suggest that a higher population in one mode leads to stronger quantum fluctuations, making it harder for the other mode to gain control and steer the state. Consequently, the criteria for quantum steering, $\mathcal{G}^{1 \to 2} > 0$ and $\mathcal{G}^{2 \to 1} > 0$, can be expressed as \cite{zheng2019manipulation}: $|\langle m_1 m_2 \rangle| > \sqrt{\langle m_2^\dagger m_2 \rangle (\langle m_1^\dagger m_1 \rangle + \frac{1}{2})}$ and $|\langle m_1 m_2 \rangle| > \sqrt{\langle m_1^\dagger m_1 \rangle (\langle m_2^\dagger m_2 \rangle + \frac{1}{2})}$, respectively.  A similar phenomenon is observed in Fig. \ref{F3}(c), where the degree of squeezing transferred from the cavity field to the magnon modes is shown. The squeezing is more pronounced for the mode with the larger coupling strength, indicating that stronger interactions lead to greater quantum state manipulation. This suggests that the mode with significant squeezing, which reflects reduced quantum uncertainty in one quadrature, is more effective in steering the other mode. This increased capability for steering arises because squeezing enhances the quantum correlations between the modes, thereby facilitating control over the quantum state of the other mode. This enhanced steering potential for squeezed states is crucial for quantum information and communication applications, where precise manipulation of quantum states is required.

	\begin{figure}[tbh!]
	\centerline{\includegraphics[width=16cm]{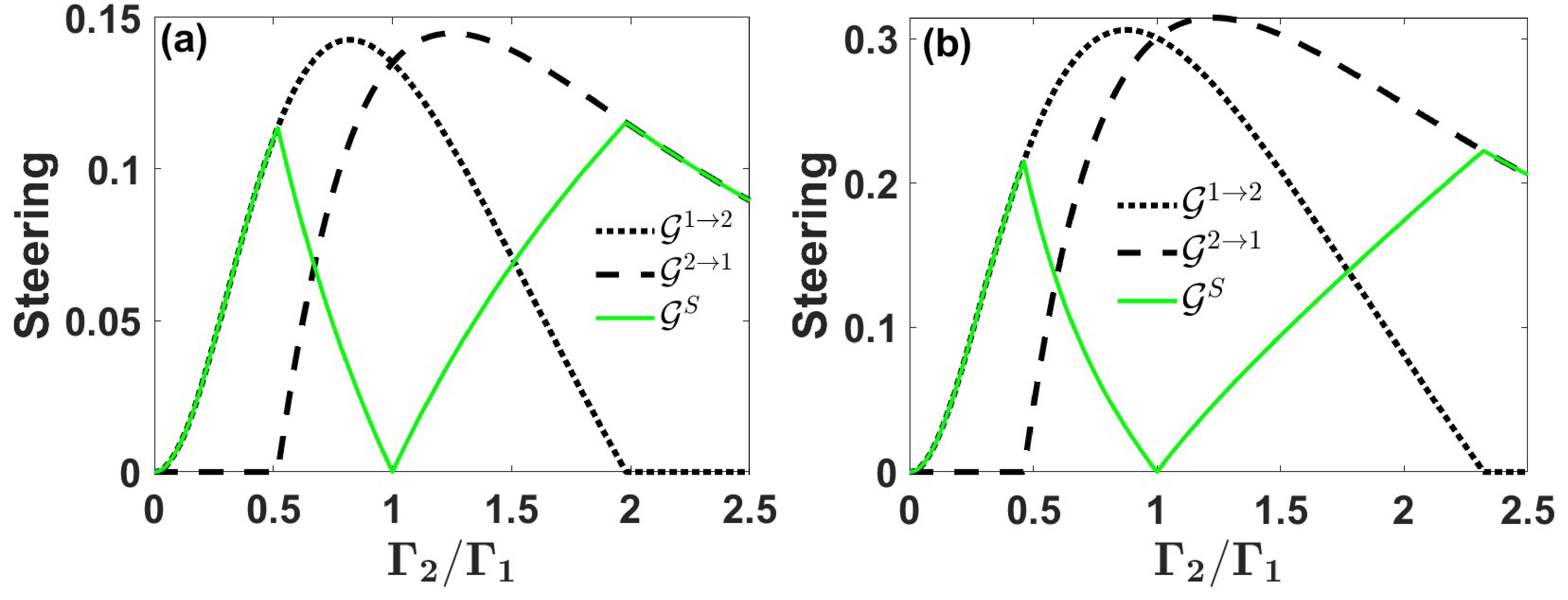}}
	\caption{\footnotesize{Plots of the degrees of steering ($\mathcal{G}^{1 \to 2}$, $\mathcal{G}^{2 \to 1}$, and $\mathcal{G}^{S}$) versus $\Gamma_{2}/\Gamma_{1}$, shown in (a) without OPA and in (b) with OPA.			
	}}
	\label{F4}
\end{figure}

In Fig. \ref{F4}, we show the steering between the two magnon modes as a function of the ratio of the coupling strengths $\Gamma_2/\Gamma_1$ under two conditions: (a) $\Lambda = 0$ (without the OPA) and (b) $\Lambda=0.49\kappa_a \neq 0$ (with the OPA). In both scenarios, the directional steering can be modulated by adjusting the coupling strengths, aligning with the findings in Fig. \ref{F3}(a). Notably, including  the OPA roughly doubles the quantum steering, as observed in Fig. \ref{F4}(b). This increase can be attributed to the OPA's ability to enhance quantum correlations  between the magnon modes, leading to a stronger manifestation of steering. Furthermore, the OPA significantly broadens the parameter space where steering directionality is maintained, indicating a more robust and versatile control over the system's quantum state. This expanded range is crucial for practical applications, as it facilitates the generation and detection of steering in experimental setups and allows for greater flexibility in tuning the system to desired quantum states.

\begin{figure}[tbh!]
	\centerline{\includegraphics[width=12cm]{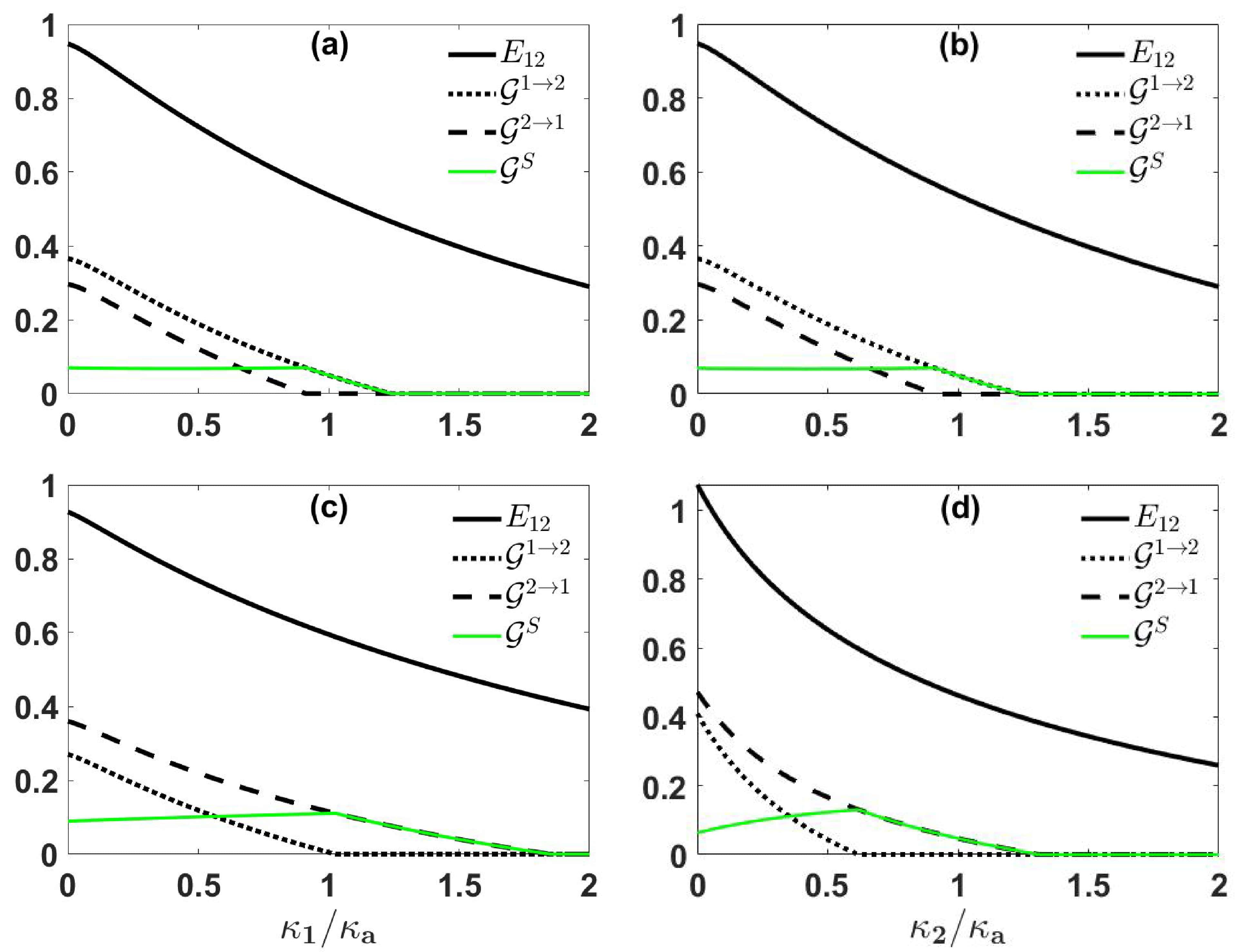}}
	\caption{\footnotesize{The entanglement  ($E_{12}$) and quantum steering as a function of (a), (c) $\kappa_1/\kappa_a$ and (b), (d) $\kappa_2/\kappa_a$. We have chosen $g_2=3\kappa_a$ in (a) and (b), and $g_2=6\kappa_a$ in (c) and (d). The other parameters are the same as in Fig. \ref{F2}.	}}
	\label{F5}
\end{figure}

In Fig. \ref{F5}, we examine how the dissipation rates of individual magnon modes affect steering and entanglement. This figure depicts the behaviour  of these quantum properties as a function of the dissipation rates $\kappa_1$ and $\kappa_2$ for two coupling strength values, $g_2 = 3\kappa_a$ and $ g_2 = 6\kappa_a$. The results show that increasing the dissipation rates of the magnon modes generally leads to a degradation of both entanglement and steering. This occurs because dissipation introduces decoherence and energy loss into the system, disrupting the coherent quantum correlations necessary for maintaining high levels of entanglement and steering. Interestingly, the reduction in entanglement and steering does not depend on the specific direction of steering; increasing $\kappa_1$ and $\kappa_2$ uniformly decreases these quantities without changing the preferred steering direction. This indicates that while dissipation affects the overall strength of quantum correlations, it does not inherently bias the system towards one steering direction. Notably, when comparing these results to the scenario where only squeezed light is present, as explored in Ref. \cite{yang2021controlling}, introducing  the OPA  increases  the directionality of quantum steering. In the presence of a squeezed field, the effect of the OPA is to amplify the quantum fluctuations and correlations within the system, thereby overcoming some of the decoherence effects introduced by dissipation. This amplification enables the system to maintain a higher degree of steering across a broader range of parameters. Specifically, the OPA increases the asymmetry in the quantum correlations between the modes, allowing for more pronounced steering in one direction even as dissipation rises. This result aligns with the findings in Fig. \ref{F4}, where the presence of the OPA was shown to expand the parameter space in which significant steering can be observed.


\begin{figure}[tbh!]
	\centerline{\includegraphics[width=14cm]{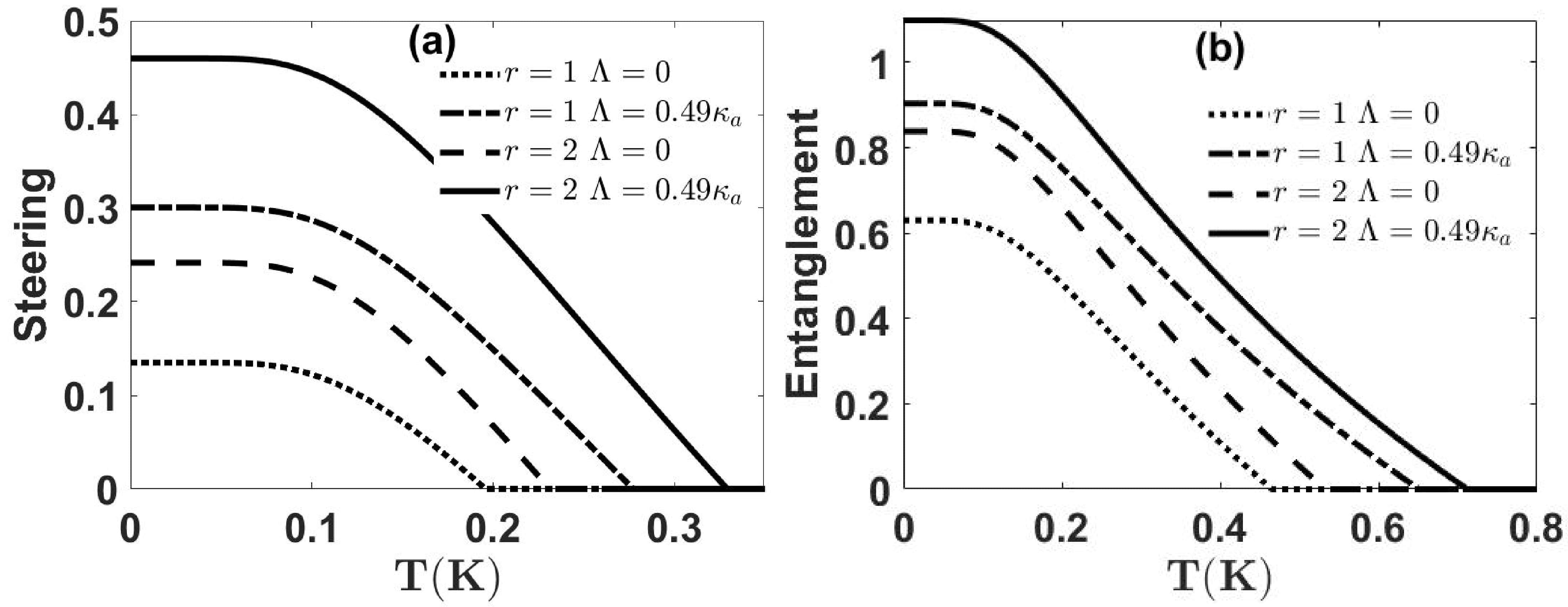}}
	\caption{\footnotesize{Plots of (a) quantum steering and (b) entanglement as a functions of the environmental temperature $T$, for different values of the squeezing parameter $r$ and the OPA gain $\Lambda$.
	}}
	\label{F6}
\end{figure}

Finally, we investigate the effect of environmental temperature $T$ on symmetric steering and entanglement by varying the gain of the OPA, $\Lambda$, and the squeezing parameter, $r$. Increasing both $r$ and $\Lambda$ enhances the robustness of steering and entanglement against temperature fluctuations. Specifically, when the squeezing parameter is $r = 2$ and the OPA gain is $\Lambda = 0.49\kappa_a$, steering and entanglement remain detectable up to temperatures of $T = 320$ mK and $T = 700$ mK, respectively. This is the best temperature range observed compared to systems without the combined effects of both  the OPA and the JPA \cite{li2019entangling,zhang2019quantum,yang2021controlling,nair2020deterministic}. The enhancement of robustness can be attributed to the role of squeezing and the OPA in generating and maintaining strong quantum correlations within the system. This investigation demonstrates that by optimizing the OPA gain $\Lambda$ and the squeezing parameter $r$, it is possible to extend the operational temperature range of quantum systems, making them more practical for real-world applications. The ability to sustain entanglement and steering under varying thermal conditions thus represents a significant advancement in  quantum information science.



	\section{Conclusion}\label{sec5}

In this paper, we have thoroughly investigated a cavity-magnon system consisting of a cavity microwave field (photons) and two magnon modes within YIG spheres, influenced by an OPA and driven by  a JPA. Our findings demonstrate that the combined presence of JPA and OPA enhances the degree of squeezing, EPR steering, and entanglement and significantly improves the system's robustness against thermal effects. This enhancement is attributed to the effective transfer of cavity field squeezing to the magnon modes via magnetic dipole interactions. Moreover, the simultaneous use of JPA and OPA broadens the  parameters under which entanglement and steering are observed, allowing for greater flexibility in experimental setups. The photon-magnon coupling strength emerges as a crucial parameter, offering precise control over the directionality of EPR steering. Our results indicate that while the steering intensity decreases with increasing dissipation rates of the individual magnon modes, the directionality remains unaffected. This stability in directional control underscores the potential for achieving one-way EPR steering by finely tuning system parameters. Our study contributes to the broader understanding of macroscopic quantum correlations, highlighting the significant role of combined parametric amplification techniques in enhancing and manipulating quantum phenomena. The insights gained from this work offer promising applications in quantum information processing, particularly in  generating, controlling, and enhancing quantum steering effects. This research lays the groundwork for future explorations into the practical implementation of cavity-magnon systems in advanced quantum technologies.


	

\end{document}